\documentclass[twocolumn]{aastex7}
\usepackage{capt-of}
\usepackage{mathtools}
\usepackage{longtable}

\shorttitle{Six New Eclipsing AM CVn Binaries from ZTF}
\shortauthors{Householder et al.}
\graphicspath{{./}{figures/}}
\usepackage{orcidlink}
\usepackage[T1]{fontenc}
\usepackage{xcolor}
\newcommand{\rev}[1]{#1}


\begin{document}

\title{Six New Eclipsing AM CVn Binaries from the Zwicky Transient Facility}
\author[0000-0002-5812-3236]{Aaron Householder}
\altaffiliation{NSF Graduate Research Fellow}
\affiliation{Department of Earth, Atmospheric and Planetary Sciences, Massachusetts Institute of Technology, Cambridge, MA 02139, USA}
\affiliation{Kavli Institute for Astrophysics and Space Research, Massachusetts Institute of Technology, Cambridge, MA 02139, USA}
\email{aaron593@mit.edu}

\author[0000-0002-7226-836X]{Kevin B. Burdge}
\affiliation{Department of Physics, Massachusetts Institute of Technology, Cambridge, MA 02139, USA}
\affiliation{Kavli Institute for Astrophysics and Space Research, Massachusetts Institute of Technology, Cambridge, MA 02139, USA}
\email{kburdge@mit.edu}

\author[0000-0003-0097-4414]{Yapeng Zhang}
\altaffiliation{51 Pegasi b Fellow}
\affiliation{Department of Astronomy, California Institute of Technology, Pasadena, CA 91125, USA}
\email{yapzhang@caltech.edu}

\author[0000-0003-4780-4105]{Emma Chickles}
\affiliation{Department of Physics, Massachusetts Institute of Technology, Cambridge, MA 02139, USA}
\affiliation{Kavli Institute for Astrophysics and Space Research, Massachusetts Institute of Technology, Cambridge, MA 02139, USA}
\email{echickle@mit.edu}

\author[0000-0002-2626-2872]{Jan van Roestel}
\affiliation{Institute of Science and Technology Austria (ISTA), Am Campus 1, 3400 Klosterneuburg, Austria}
\email{j.c.j.vanroestel@uva.nl}

\author[0000-0002-0948-4801]{Matthew J. Green}
\affiliation{Homer L. Dodge Department of Physics and Astronomy, University of Oklahoma, 440 W. Brooks Street, Norman, OK 73019, USA}
\affiliation{JILA, University of Colorado and National Institute of Standards and Technology, 440 UCB, Boulder, CO 80309-0440, USA}
\email{mjgreenastro@gmail.com}

\author[0000-0002-8810-858X]{Kaley Brauer}
\affiliation{Center for Astrophysics \textbar\ Harvard \& Smithsonian, 60 Garden Street, Cambridge, MA 02138, USA}
\email{kaley.brauer@cfa.harvard.edu}

\author[0000-0002-0568-6000]{Joheen Chakraborty}
\affiliation{Department of Physics, Massachusetts Institute of Technology, Cambridge, MA 02139, USA}
\affiliation{Kavli Institute for Astrophysics and Space Research, Massachusetts Institute of Technology, Cambridge, MA 02139, USA}
\email{joheen@mit.edu}

\author[0000-0002-6401-778X]{Jack Dinsmore}
\affiliation{Department of Physics, Stanford University, Stanford, CA 94305, USA}
\affiliation{Kavli Institute for Particle Astrophysics and Cosmology, Stanford University, Stanford, CA 94305, USA}
\email{jtd@stanford.edu}

\author[0000-0001-8467-9767]{Gabor Furesz}
\affiliation{Kavli Institute for Astrophysics and Space Research, Massachusetts Institute of Technology, Cambridge, MA 02139, USA}
\email{gfuresz@mit.edu}

\author[0000-0003-1361-985X]{Juliana Garc\'ia-Mej\'ia}
\altaffiliation{51 Pegasi B Fellow, MIT Pappalardo Physics Fellow}
\affiliation{Kavli Institute for Astrophysics and Space Research, Massachusetts Institute of Technology, Cambridge, MA 02139, USA}
\affiliation{Center for Astrophysics \textbar\ Harvard \& Smithsonian, 60 Garden Street, Cambridge, MA 02138, USA}
\email{jgarciam@mit.edu}

\author[0000-0002-3168-0139]{Matthew J. Graham}
\affiliation{Division of Physics, Mathematics, and Astronomy, California Institute of Technology, Pasadena, CA 91125, USA}
\email{mjg@caltech.edu}

\author[0000-0002-3239-5989]{Aurora Y. Kesseli}
\affiliation{IPAC, Mail Code 100-22, Caltech, 1200 E. California Boulevard, Pasadena, CA 91125, USA}
\email{aurorak@ipac.caltech.edu}

\author[0000-0001-5390-8563]{Shrinivas R. Kulkarni}
\affiliation{Division of Physics, Mathematics, and Astronomy, California Institute of Technology, Pasadena, CA 91125, USA}
\email{srk@astro.caltech.edu}

\author[0000-0002-7191-4403]{Christopher Layden}
\affiliation{Department of Physics, Massachusetts Institute of Technology, Cambridge, MA 02139, USA}
\affiliation{Kavli Institute for Astrophysics and Space Research, Massachusetts Institute of Technology, Cambridge, MA 02139, USA}
\email{clayden7@mit.edu}

\author[0000-0001-6331-112X]{Geoffrey Mo}
\email{gmo@caltech.edu}
\affiliation{Department of Astronomy, California Institute of Technology, Pasadena, CA 91125, USA}
\affiliation{The Observatories of the Carnegie Institution for Science, Pasadena, CA 91101, USA}

\author[0000-0003-0412-9664]{David Osip}
\email{dosip@carnegiescience.edu}
\affiliation{Las Campanas Observatory, Carnegie Institution for Science, Colina el Pino, Casilla 601 La Serena, Chile}

\author[0000-0001-6711-3286]{Roger W. Romani}
\affiliation{Department of Physics, Stanford University, Stanford, CA 94305, USA}
\affiliation{Kavli Institute for Particle Astrophysics and Cosmology, Stanford University, Stanford, CA 94305, USA}
\email{rwr@stanford.edu}

\author[0000-0001-7246-5438]{Andrew Vanderburg}
\affiliation{Center for Astrophysics \textbar\ Harvard \& Smithsonian, 60 Garden Street, Cambridge, MA 02138, USA}
\email{avanderburg@cfa.harvard.edu}

\begin{abstract}
We present the discovery of six new eclipsing AM CVn binaries, helium-accreting ultracompact systems in which a cool donor eclipses the accreting white dwarf. The systems were found in a GPU-accelerated Box Least Squares (BLS) search of Zwicky Transient Facility (ZTF) photometry and have orbital periods of 39.3 to 59.9 minutes. We obtained spectroscopy with the Low Resolution Imaging Spectrometer (LRIS) on the Keck I telescope for all six systems, and every spectrum shows helium emission with no detectable hydrogen, confirming their AM CVn classification. Our discoveries increase the number of deeply eclipsing AM CVn from 11 to 17 (i.e., systems in which the accreting white dwarf is both the dominant source of optical light and eclipsed by the donor). With high-speed photometric follow-up, these systems will provide the mass ratios, inclinations, and fractional radii of both components. They will also provide new tests of AM CVn formation models and stronger constraints on the long-term evolution of these sources. 
\end{abstract}

\section{Introduction}\label{sec:intro}

AM CVn binaries are ultracompact mass-transferring systems, with orbital periods between roughly 5 and 70 minutes, in which a white dwarf accretes helium-rich material from a degenerate or semi-degenerate donor star \citep{2010PASP..122.1133S, 2025AandA...700A.107G}. Eclipsing systems are particularly valuable because the eclipse geometry constrains the component masses and radii, which are difficult to obtain from spectroscopy alone \citep{2018MNRAS.476.1663G, 2022MNRAS.512.5440V}. In these systems, the donor eclipses the accreting white dwarf and its surrounding accretion disk. At the longer orbital periods, where the mass-transfer rate is low and the disk is faint, the accreting white dwarf dominates the optical light, so the eclipse is primarily an eclipse of the white dwarf itself. \rev{When the light curve shows eclipses of both the white dwarf and bright spot, photometry yields the mass ratio, inclination, and fractional radii \citep{1984MNRAS.207..705C, 1986MNRAS.219..629W, 2011MNRAS.410.1113C, 2018MNRAS.476.1663G}. Though we note that absolute masses and radii must use an accretor mass-radius relation, which is somewhat uncertain for accreting systems like these. Nonetheless, these} measurements test \rev{the white dwarf donor, helium star donor, and evolved cataclysmic variable (CV) channels}, which predict different donor mass-radius relationships \citep{2001AandA...368..939N, 2003MNRAS.340.1214P}.

Prior to the Zwicky Transient Facility \citep[ZTF;][]{2019PASP..131a8002B, 2019PASP..131g8001G, 2019PASP..131a8003M}, only two AM CVn systems were known to show eclipses of the central white dwarf: YZ LMi \citep{2005AJ....130.2230A, 2011MNRAS.410.1113C} and Gaia14aae \citep{2015MNRAS.452.1060C, 2018MNRAS.476.1663G}. The high cadence and wide-field coverage of ZTF have since made systematic searches for eclipsing AM CVn possible. \citet{2022MNRAS.512.5440V} discovered five new eclipsing AM CVn systems with orbital periods of 35.4 to 61.5 minutes by searching for eclipses in the ZTF light curves of Gaia-selected white dwarfs. Four more were found in ZTF and SRG/eROSITA data \citep{2021AJ....162..113V, 2023ApJ...954...63R, 2024AandA...683L..10K}. In total, the \citet{2025AandA...700A.107G} catalog lists 11 \rev{AM\,CVn binaries in which the accreting white dwarf dominates the optical light and is eclipsed by the donor}. Additional ultracompacts at shorter orbital periods also show eclipses, but their higher mass-transfer rates cause the accretion disk and stream-impact region to dominate the optical emission \rev{\citep[e.g.,][]{2014ApJ...785..114L, 2021AandA...645A.114B, 2024ApJ...977..262C, 2026ApJ..1000..237C}.} \rev{Any occultation of the accreting white dwarf then contributes little to the observed eclipse and provides little constraint on its radius.} In this \rev{paper}, we \rev{discuss} the systems in which the accreting white dwarf dominates the optical light and is eclipsed by the donor, so that the eclipse constrains the \rev{fractional} radii of both components. \rev{In particular, we present six new eclipsing AM CVn binaries from a systematic Box Least Squares (BLS) search of ZTF photometry (Householder et al., in prep).} We describe the discovery of these systems in ZTF in Section~\ref{sec:discovery}, present the follow-up spectroscopy that confirms them as AM CVn binaries in Section~\ref{sec:spectroscopy}, discuss the implications for the eclipsing AM CVn population in Section~\ref{sec:discussion}, and summarize our findings in Section~\ref{sec:conclusion}.

\section{Discovery in ZTF}\label{sec:discovery}

\rev{\subsection{ZTF Photometry and BLS Search}}

\rev{We found these six systems in a systematic search of ZTF data (Householder et al., in prep).} ZTF uses a 47 deg$^2$ camera on the 48-inch Samuel Oschin Schmidt Telescope at Palomar \rev{and images} the sky every few days with \rev{30\,s} exposures \citep{2019PASP..131a8002B, 2019PASP..131g8001G, 2019PASP..131a8003M}. \rev{These observations allowed us to perform forced PSF photometry on ZTF difference images at fixed positions of about 1.3 billion Pan-STARRS1 DR1 sources with more than five detections \citep{2020ApJS..251....7F, 2019PASP..131a8003M}. We retained sources with at least 50 measurements and concatenated median-shifted $g$ and $r$ light curves. }\rev{Each Pan-STARRS1 DR1 position is the mean of observations from 2010 to 2014 \citep{2016arXiv161205560C}, about 10 years before the mean ZTF epoch. }\rev{We did not propagate these positions for proper motion, so our search may miss some nearby high proper motion systems (see discussion in Section~\ref{sec:candidates}).} \rev{After building these light curves, we ran a GPU-accelerated BLS search \citep{2002AandA...391..369K} for each source over trial periods as short as 4 minutes, using roughly $10^7$ trial periods per source.}

\subsection{\rev{Candidate Selection}}\label{sec:candidates}

\rev{The BLS search produced many short-period eclipsing candidates, which we ranked by eclipse significance and visually inspected before prioritizing six systems for spectroscopy based on eclipse significance and observability. ZTF\,J1302$-$1746 and ZTF\,J0859$-$1110 were also found independently by an unpublished search (van Roestel et al., in prep) that targeted white dwarfs pre-selected from Gaia. Our search instead covered about 1.3 billion Pan-STARRS1 DR1 sources without white dwarf pre-selection. Our search also found 20 additional eclipsing candidates with periods below 70 minutes that are not presented here and need more follow-up. Table~\ref{tab:known} gives the coordinates, periods, and photometric properties of the six new systems together with the 11 previously known eclipsing systems.}

\rev{To test how well our search recovered known eclipsing AM CVn binaries, we checked the 11 previously known systems and recovered 9 at high significance at their published periods. YZ LMi \citep{2005AJ....130.2230A, 2011MNRAS.410.1113C} had its strongest periodogram peak at the correct period but at very low significance, so it would not have been selected among our candidates. SRGe J0453+6224 \citep{2023ApJ...954...63R} was not recovered because it only had 215 epochs that we reduced, making it challenging to detect (all other sources had more than 400 epochs). We therefore acknowledge that our search likely does not recover every AM CVn binary that is detectable in ZTF. Several selection effects further complicate the recovered sample. First, we did not propagate the 2010-2014 mean Pan-STARRS1 DR1 positions for proper motion, so high proper motion systems could be missed. ZTF cadence also varies from field to field, so the number of epochs available for recovery differs across the sky. Short period AM CVn binaries can also outburst frequently, while systems above about 50 min rarely do \citep{2015MNRAS.446..391L, 2025AandA...700A.107G}. These outbursts can dominate periodograms and act as catastrophic outliers in a Box Least Squares search. Furthermore, AM CVns are expected to become dimmer as the period grows (see Figure~\ref{fig:mgporb}), so a magnitude limited survey contains fewer long period systems. Together, these effects mean we caution against treating Table~\ref{tab:known} as complete or representative of the true period distribution. The 20 additional candidate systems mentioned above also still need follow-up, so a separate paper will quantify completeness with comprehensive injection and recovery simulations (Chakraborty et al., in prep).}

\begin{figure*}[ht]
    \centering
    \includegraphics[width=0.85\linewidth]{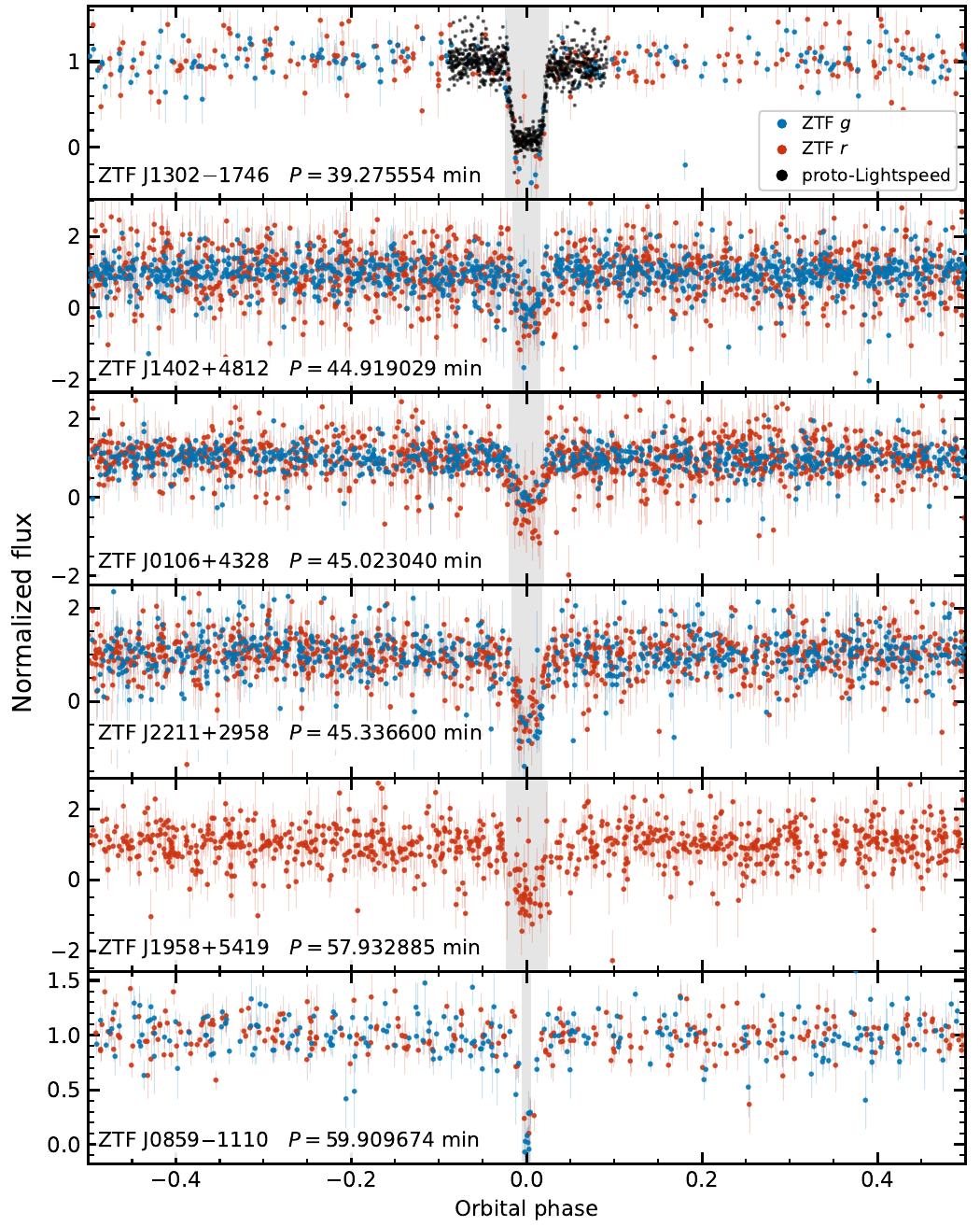}
    \caption{Phase-folded ZTF light curves of the six new eclipsing AM CVn systems, ordered by orbital period. Blue and red points are individual $g$- and $r$-band exposures, and the gray band in each panel marks the eclipse duration from the BLS fit, or from the fit to the proto-Lightspeed data for ZTF\,J1302$-$1746. The black points in the ZTF\,J1302$-$1746 panel are unfiltered 1\,s proto-Lightspeed exposures \citep{2026PASP..138f5001L, 2026SPIE14149E..2IL} of two consecutive eclipses, folded on the same ephemeris. Only the portion of the proto-Lightspeed light curve around the eclipses is shown due to poor weather outside of the eclipse. For ZTF\,J1958+5419 only $r$-band data are shown because of an unreliable $g$-band reference flux.}
    \label{fig:lightcurves}
\end{figure*}

\rev{\subsection{Eclipse Measurements}}

\rev{Figure~\ref{fig:lightcurves} shows the phase-folded ZTF light curves, each with} an eclipse on a relatively flat out-of-eclipse \rev{baseline. For} \rev{five of the six systems, BLS fits at the recovered periods give} eclipse durations of roughly 0.7 to 2.8 \rev{minutes (Table~\ref{tab:known}, gray bands in Figure~\ref{fig:lightcurves}).} Durations this short are consistent with the donor eclipsing the accreting white dwarf, which dominates the optical light at these long orbital periods. \rev{The near-zero in-eclipse fluxes of all six systems, together with their 39 to 60 minute periods where the disk is faint, indicate that the accreting white dwarf supplies most of the optical light. Importantly, the 30\,s} exposures do not resolve ingress and egress, so the \rev{durations are rough estimates. We also note that some in-eclipse points for the faintest systems} fall below zero flux, which is unphysical and likely reflects an unreliable reference flux, so we do not quote eclipse depths. \rev{Measuring the} contact points and depths \rev{of these systems} will need high-speed photometry. \rev{As a first step, we observed ZTF\,J1302$-$1746 with proto-Lightspeed, a high-speed imager on the Magellan Clay telescope \citep{2026PASP..138f5001L, 2026SPIE14149E..2IL}, on 2026 May 24 UT. }\rev{The observation covered} just under two orbital periods in continuous 1\,s exposures \rev{without a filter, during poor weather outside eclipse.} \rev{We reduced these data} with the aperture photometry pipeline used in prior proto-Lightspeed work \citep{2026PASP..138f5001L, 2026ApJ..1000..237C}. We fit the two eclipses with a simple \texttt{ellc} \citep{2016AandA...591A.111M} model of a white dwarf eclipsed by a dark Roche-lobe-filling \rev{donor. }\rev{The fit provided the mid-eclipse time and duration listed in Table~\ref{tab:known}, and we leave detailed modeling to a follow-up paper.}

\section{Keck/LRIS Spectroscopy}\label{sec:spectroscopy}

\begin{figure*}[t]
    \centering
    \includegraphics[width=0.78\linewidth]{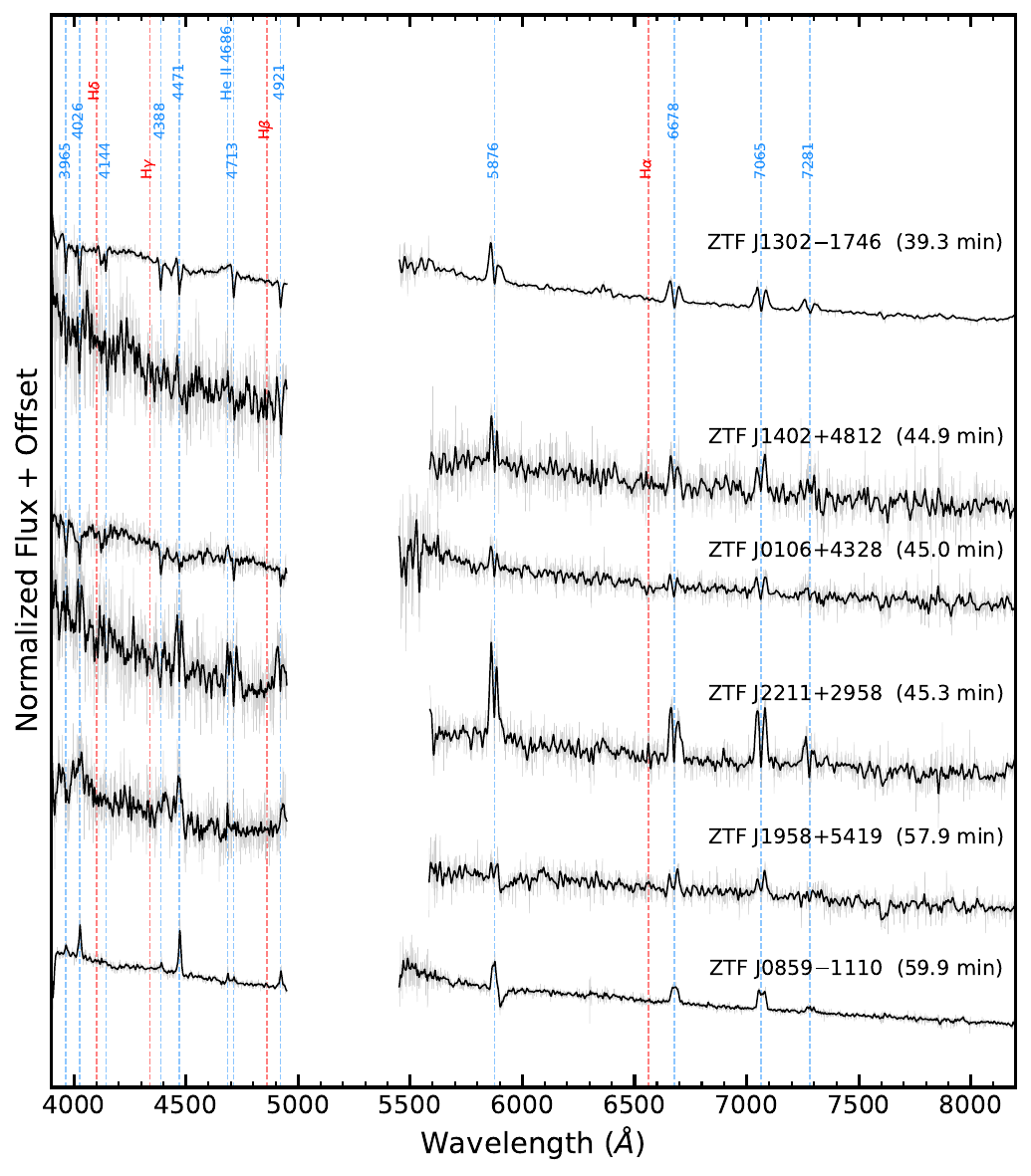}
    \caption{Keck/LRIS spectra of the six new eclipsing AM CVn systems, ordered by orbital period. The unsmoothed spectrum is shown in gray with a lightly smoothed version in black, each scaled by its median flux and offset vertically. He\,\textsc{i} lines and He\,\textsc{ii} 4686\,\AA\ are marked and labeled with dashed blue lines, and the Balmer lines, which are not detected in any system, with dashed red lines. In ZTF\,J1302$-$1746 and ZTF\,J0106+4328 the blue-arm He\,\textsc{i} lines are in absorption, which has been seen in other eclipsing systems \citep[e.g.,][]{2022MNRAS.512.5440V}, and ZTF\,J1302$-$1746 shows weak signs of Si\,\textsc{ii} 6347 and 6371\,\AA, but more spectroscopy is needed to confirm them. The region near 5000--5450\,\AA, where the throughput of both arms drops toward the D560 dichroic crossover, is not shown.}
    \label{fig:spectra}
\end{figure*}

To confirm the AM CVn classification of the candidates, we obtained optical spectroscopy with the Low Resolution Imaging Spectrometer \citep[LRIS;][]{1995PASP..107..375O} on the Keck I 10-m telescope. We used the 600/4000 grism on the blue arm \citep[LRIS-B;][]{2004ApJ...604..534S} and the 600/7500 and 400/8500 gratings on the red arm \citep{2010SPIE.7735E..0RR} with the D560 dichroic and the 1\arcsec\ longslit, with exposure times of 300 to 900\,s. The spectra were reduced using the \texttt{LPipe} pipeline\footnote{\url{https://sites.astro.caltech.edu/~dperley/programs/lpipe.html}} \citep{2019PASP..131h4503P}, with flux calibration from observations of spectrophotometric standards obtained on the same night with the same instrument setup. The reduced spectra (Figure~\ref{fig:spectra}) show prominent He\,\textsc{i} lines, including He\,\textsc{i} 4471, 4921, 5876, 6678, 7065, and 7281\,\AA, with no detectable Balmer series or other hydrogen features. The absence of hydrogen together with the strong helium emission thus confirms that all six systems are AM CVn-type binaries.

In \rev{most} systems, the helium emission lines are broad \rev{and often} double-peaked, most prominently in ZTF\,J2211+2958 (Figure~\ref{fig:hezoom}). Double-peaked emission is the signature of a rotating accretion disk, with the two peaks coming from the approaching and receding halves of the disk \citep{1981AcA....31..395S, 1986MNRAS.218..761H}. ZTF\,J0859$-$1110, the longest-period system in our sample, instead has lines that are narrower and predominantly single-peaked. Single-peaked emission would normally suggest a low inclination, which the eclipses rule out. Instead, the narrow single-peaked lines of ZTF\,J0859$-$1110 may be analogous to the ``central spike'' seen in GP\,Com and V396\,Hya, two systems with long orbital periods and low accretion rates, in which a narrow component moves with the accreting white dwarf and is attributed to emission from near the white dwarf surface \citep{1999MNRAS.304..443M, 2016MNRAS.457.1828K}. Phase-resolved spectroscopy of ZTF\,J0859$-$1110 could test this interpretation.

\rev{While most of our systems show He\,\textsc{i} emission, ZTF\,J1302$-$1746 (39.3 minutes) and ZTF\,J0106+4328 (45.0 minutes) instead show absorption at the centers of the He\,\textsc{i} lines blueward of 5100\,\AA. Such absorption also appears to be common among eclipsing systems, as \citet{2022MNRAS.512.5440V} found He\,\textsc{i} lines in both double-peaked emission and absorption in all five of their eclipsing systems. One possible explanation is absorption in the photosphere of the accreting white dwarf, which supplies most of the optical light at these periods. This interpretation has precedent in GP Com, where \citet{2016MNRAS.457.1828K} attributed absorption lines to the white dwarf photosphere because their radial velocity amplitude, phase and mean velocity matched the white dwarf. Alternatively, an edge-on optically thick accretion disk could produce absorption at line center because the contrast between the disk continuum and optically thick line emission decreases toward bluer wavelengths. Both explanations remain viable for ZTF\,J1302$-$1746 and ZTF\,J0106+4328, but phase-resolved spectroscopy could distinguish them by measuring whether the absorption depth varies with orbital phase. Beyond the He\,\textsc{i} absorption, ZTF\,J1302$-$1746 also shows weak signs of Si\,\textsc{ii} 6347 and 6371\,\AA, similar to lines seen in SDSS J1240$-$0159 \citep{2005MNRAS.361..487R}, SDSS J0129+3842 and SDSS J1525+3600 \citep{2013MNRAS.432.2048K}, and three eclipsing systems \citep{2022MNRAS.512.5440V}. More spectroscopy is needed to confirm these features in ZTF\,J1302$-$1746.}

\section{Discussion}\label{sec:discussion}

\subsection{Growth of the Eclipsing AM CVn Population}

Before this work, 11 AM CVn binaries were known in which the accreting white dwarf dominates the optical light and is eclipsed \citep{2025AandA...700A.107G}, and the six systems presented here increase the sample to 17 \rev{(with 20 additional eclipsing candidates that still need follow-up)}. As an AM CVn binary ages, the donor becomes increasingly degenerate \citep{2007MNRAS.381..525D}, the orbit widens, and the mass-transfer rate falls, so systems spend most of their lifetime at long periods \citep{2001AandA...368..939N}. The six new systems and six of the 11 known systems have periods above 39 minutes, as might be expected from this evolution. \rev{However, outbursts, proper motion, cadence, and visual vetting all complicate completeness, so this period distribution describes the systems recovered by the search rather than the underlying population. A separate paper on ultracompact binaries in ZTF will quantify completeness with comprehensive injection and recovery simulations (Chakraborty et al., in prep).}

\rev{Although we caution against treating Table~\ref{tab:known} as complete, the confirmed eclipsing systems still give a lower limit on the local space density of AM CVn binaries. In particular, five of the 17 known eclipsing AM CVn binaries have Gaia DR3 distances below 300 pc and parallax over error above 4, allowing an estimate after correcting for eclipse probability. Assuming a $0.6\,M_\odot$ accretor with $R_1=0.013\,R_\odot$ and a Roche-lobe-filling $0.02\,M_\odot$ donor, the Roche-lobe relation in Section~\ref{sec:minmass} and Kepler's third law give a donor radius of about $0.06\,R_\odot$. For these five systems, with orbital periods from 49.7 to 61.5 minutes, the corresponding eclipse probabilities $(R_1+R_2)/a$ range from 0.178 to 0.183 ($\sim 0.18$). Correcting for the eclipse probability of about 0.18 gives a lower limit of $2.5\times10^{-7}$ pc$^{-3}$, or 28 AM CVn binaries within 300 pc. This estimate is similar to the $3\times10^{-7}$ pc$^{-3}$ lower limit from \citet{2018AandA...620A.141R} and consistent at 1 sigma with the $(5\pm3)\times10^{-7}$ pc$^{-3}$ estimate from \citet{2013MNRAS.429.2143C}. However, the calculation assumes component properties that may not hold for these systems, so injection and recovery simulations together with improved modeling will be needed to measure the space density more accurately.}

\subsection{\rev{Implications for Formation and Evolutionary Channels}}\label{sec:channels}

AM CVn binaries are thought to form through three main evolutionary channels: the white dwarf donor channel, the helium star donor channel, and the evolved cataclysmic variable (CV) channel \citep{2001AandA...368..939N, 2003MNRAS.340.1214P}. These channels predict different relationships between donor mass, radius, and orbital period, so eclipse light curve modeling of the systems reported here will constrain these quantities and provide new tests of the formation models. In particular, the growing sample of eclipsing systems at long orbital periods ($P_{\rm orb} \gtrsim 40$ minutes) will help distinguish the white dwarf donor and evolved CV channels, which predict different donor properties at these periods \citep{2022MNRAS.512.5440V}. Beyond the mass-radius relationship, the composition of the accreted material also depends on the formation channel. For instance, \citet{2010MNRAS.401.1347N} showed that the nitrogen-to-carbon (N/C) abundance ratio in optical and ultraviolet spectra can help distinguish white dwarf donors from helium star donors. Deeper spectra to measure the donor compositions of these systems \rev{\citep[see e.g.,][]{2013MNRAS.432.2048K}} will therefore provide a second test of their origin.

\begin{figure*}[t]
    \centering
    \includegraphics[width=0.98\textwidth]{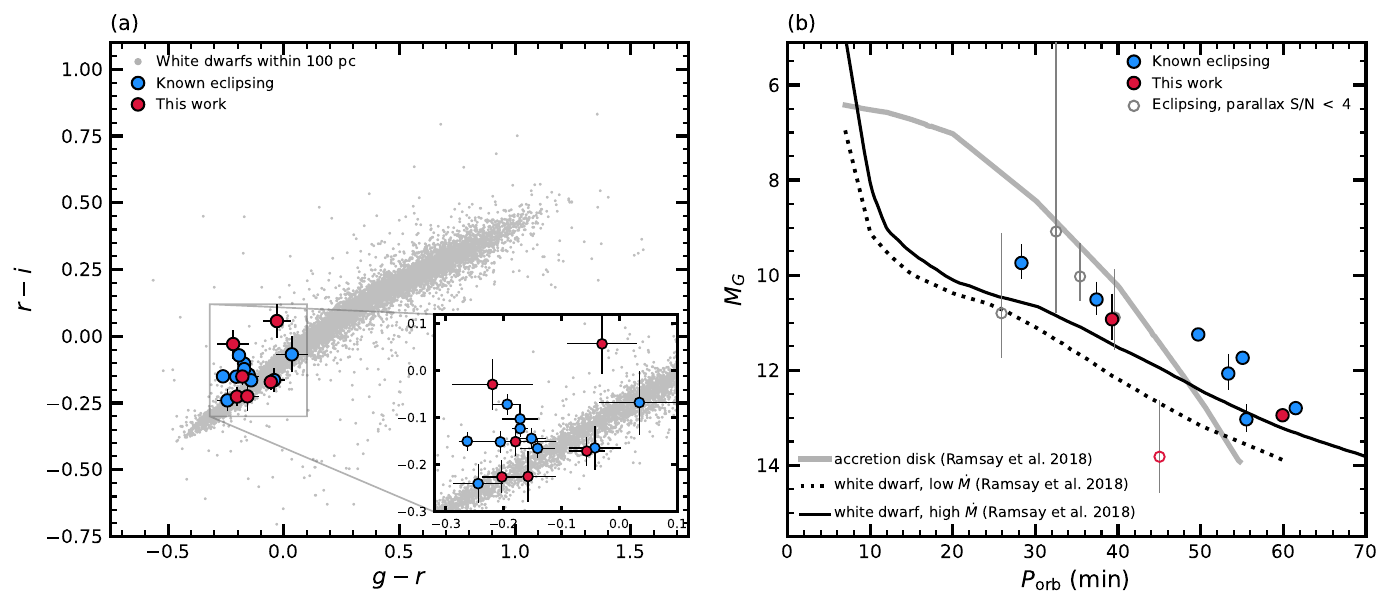}
    \caption{\rev{(a) Pan-STARRS DR1 mean PSF $g-r$ and $r-i$ colors for Gaia white dwarfs within 100 pc \citep{2021MNRAS.508.3877G} and eclipsing AM CVn binaries. (b) Gaia DR3 absolute G magnitude versus orbital period, with the disk and low and high mass-transfer rate relations from \citet{2018AandA...620A.141R}. In both panels, filled blue circles mark previously known eclipsing systems and filled red circles mark the new systems, but ZTF\,J0729$-$0602 is omitted from panel (a) because its Pan-STARRS DR1 mean PSF colors are contaminated by outbursts. ZTF\,J1402+4812, ZTF\,J2211+2958 and ZTF\,J1958+5419 are absent from panel (b) because they lack Gaia parallaxes. In (b), open circles mark eclipsing systems with parallax over error below 4.}}
    \label{fig:mgporb}
\end{figure*}

\rev{The same long-period eclipsing systems also test a trend noted by \citet{2018AandA...620A.141R}, in which AM CVn binaries above about 50 minutes appear brighter than expected for their low mass-transfer rates. In Figure~\ref{fig:mgporb}(a), the eclipsing systems lie near the blue end of the Gaia white dwarf locus within 100 pc, and the inset zooms on them. To compare our systems with this trend, we show in Figure~\ref{fig:mgporb}(b) three relations digitized from Figure 3 of \citet{2018AandA...620A.141R}, including the disk relation of \citet{2004MNRAS.349..181N} and low and high mass-transfer rate relations of \citet{2006ApJ...640..466B}. Of our new systems above 50 minutes, only ZTF\,J0859$-$1110 (59.9 min) has a usable Gaia parallax, and it follows the trend at $\sim$0.9 mag above the low mass-transfer rate relation. Several explanations could account for this excess brightness, including the higher donor masses or higher accretion rates proposed by \citet{2018AandA...620A.141R}. A cooling delay in the accretor, together with uncertainties in donor initial entropy and cooling, may also explain the discrepancy according to \citet{2021ApJ...923..125W}. Ultimately, more theoretical work is likely needed to clarify how these explanations contribute. Better parallaxes would test the trend further, and Gaia DR4 may improve the ZTF\,J0106+4328 parallax and detect ZTF\,J1402+4812, ZTF\,J2211+2958 and ZTF\,J1958+5419.}

\subsection{Minimum Donor Masses}\label{sec:minmass}
\begin{figure}[ht]
    \centering
    \includegraphics[width=\linewidth]{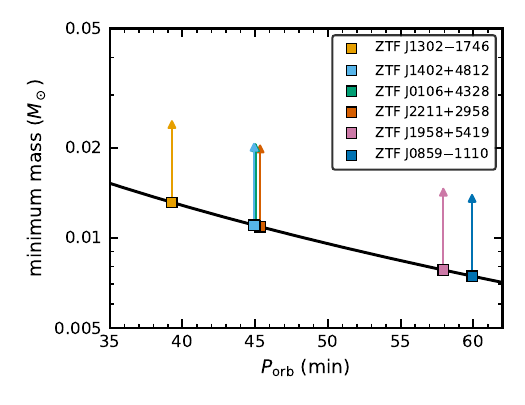}
    \caption{Minimum donor mass as a function of orbital period for a zero-temperature white dwarf donor that fills its Roche lobe (Equations~\ref{eq:eggleton} and \ref{eq:coldwd}). \rev{The upward arrows indicate that these masses are lower limits because real donors are expected to be hotter and larger than a cold white dwarf.}}
    \label{fig:minmass}
\end{figure}

Although the component masses of these systems must ultimately come from eclipse light curve modeling, the orbital period alone already places a lower limit on the donor mass (see e.g., \citealt{2024ApJ...977..262C}). Because the donor is transferring mass, we assume that it fills its Roche lobe, whose volume-equivalent radius is well approximated by
\begin{equation}
\frac{R_{\rm L}}{a} = \frac{0.49\,q^{2/3}}{0.6\,q^{2/3} + \ln\left(1 + q^{1/3}\right)},
\label{eq:eggleton}
\end{equation}
where $q = M_2/M_1$ is the mass ratio of the donor to the accretor and $a$ is the orbital separation  \citep{1983ApJ...268..368E}. Combining Equation \ref{eq:eggleton} with Kepler's third law, the mean density of a Roche-lobe-filling donor, $\bar{\rho}_2 = 3M_2/(4\pi R_{\rm L}^3)$, depends almost entirely on the orbital period and only weakly on $q$. In particular, over the mass ratios relevant to AM CVn binaries ($0.01 \lesssim q \lesssim 1$) this is well approximated by $P_{\rm orb} \approx 6.19\,{\rm min}\,(\bar{\rho}_2 / 10^4\,{\rm g\,cm^{-3}})^{-1/2}$ \citep{2024ApJ...977..262C}, so the donors in our six systems must have mean densities of roughly 100 to 250\,g\,cm$^{-3}$. A mean density alone does not determine a mass, so to obtain a minimum mass we adopt the zero-temperature white dwarf mass-radius relation of \citet{1988ApJ...332..193V},
\begin{multline}
\frac{R}{R_\odot} = 0.0114 \left[\left(\frac{M}{M_{\rm Ch}}\right)^{-2/3} - \left(\frac{M}{M_{\rm Ch}}\right)^{2/3}\right]^{1/2} \\
\times \left[1 + 3.5\left(\frac{M}{M_{\rm p}}\right)^{-2/3} + \left(\frac{M}{M_{\rm p}}\right)^{-1}\right]^{-2/3},
\label{eq:coldwd}
\end{multline}
where $M_{\rm Ch} = 1.44\,M_\odot$ and $M_{\rm p} = 0.00057\,M_\odot$. Importantly, a cold degenerate object is the most compact state available for a given mass, and its mean density increases monotonically with mass, so at each orbital period there is exactly one mass for which a cold white dwarf reaches the Roche-lobe density. Solving $\bar{\rho}_{\rm cold}(M_2) = \bar{\rho}_2(P_{\rm orb})$ therefore yields minimum donor masses between $0.007$ and $0.013\,M_\odot$ for our systems (Table~\ref{tab:known} and Figure~\ref{fig:minmass}).

Because real donors are expected to be hotter and larger than cold white dwarfs, the above minimum masses are simply lower limits, and future eclipse modeling will replace them with measurements. In particular, high-cadence light curves of these systems will resolve the eclipse ingress and egress profiles, which will allow us to search for the disk and bright-spot features seen in other eclipsing AM CVn. The eclipse of the bright spot, where the accretion stream strikes the disk, fixes the mass ratio from the light curve alone, because the eclipse of the white dwarf by itself constrains only a combination of mass ratio and inclination \citep{2011MNRAS.410.1113C, 2018MNRAS.476.1663G}. We searched the proto-Lightspeed light curve of ZTF\,J1302$-$1746 for a bright-spot hump, but the data were obtained in poor weather and do not constrain it, so a quantitative search for disk features will accompany the eclipse modeling in a follow-up paper.

\subsection{Orbital Period Derivatives as Evolutionary Tracers}

The long orbital periods of the systems reported here also make their orbital period derivatives ($\dot{P}$) valuable tests of the evolutionary models. In a mass-transferring AM CVn binary the orbit is expected to widen, at a rate set by gravitational-wave angular momentum loss and the response of the donor, so $\dot{P}$ constrains the mass-transfer rate and tests the formation and evolutionary models in a way that eclipse light curves alone cannot. Measuring $\dot{P}$ needs long baselines of precise eclipse timing, and so far only two eclipsing AM CVn systems with orbital periods longer than 20 minutes (YZ\,LMi and Gaia14aae) have been monitored long enough for such a measurement, with period derivatives consistent with zero in both cases \citep[$3\sigma$ upper limits of $1.1\times10^{-13}$ and $9.8\times10^{-14}$\,s\,s$^{-1}$;][]{2026MNRAS.548ag673G}. With angular momentum loss from gravitational waves alone, the predicted period derivatives are $+1.8\times10^{-13}$\,s\,s$^{-1}$ for YZ\,LMi and $+5.2\times10^{-14}$\,s\,s$^{-1}$ for Gaia14aae. For YZ\,LMi, the measured value of $(3.6 \pm 2.5)\times10^{-14}$\,s\,s$^{-1}$ lies more than $5\sigma$ below this prediction. \rev{The six new long-period eclipsing systems offer additional targets for future period derivative measurements, with the mid-eclipse epochs listed in Table~\ref{tab:known} serving as reference points for long-baseline precise eclipse timing.}

\begin{table*}[p]
\centering
\caption{All known AM CVn binaries in which the accreting white dwarf dominates the optical light and is eclipsed by the donor. Shorter-period eclipsing systems dominated by the accretion disk \citep[e.g.,][]{2024ApJ...977..262C, 2026ApJ..1000..237C} are not included. Magnitudes are Pan-STARRS DR1 mean PSF values, and periods \rev{and mid-eclipse epochs} of the previously known systems are published values \rev{where available. Blank epoch entries indicate that no published mid-eclipse epoch is available}. For the six new systems, the period, mid-eclipse epoch $T_0$ (MBJD$_{\rm TDB}$ = BJD$_{\rm TDB}$ $-$ 2400000.5), and approximate eclipse duration $t_{\rm ecl}$ are measured here (Section~\ref{sec:discovery}), with uncertainties on the final digits of $P_{\rm orb}$ and $T_0$ in parentheses. For ZTF\,J1302$-$1746, $T_0$ is the mid-eclipse time of the first proto-Lightspeed eclipse and $t_{\rm ecl}$ is the first-to-fourth-contact duration from the fit to those data (Section~\ref{sec:discovery}). For the other five systems, $t_{\rm ecl}$ is the box width from the BLS fit to the 30\,s ZTF exposures, which do not resolve ingress and egress, so these durations are rough estimates and tend to be underestimated. \rev{For SRGe\,J0453+6224, we note that \citet{2023ApJ...954...63R} do not state the time standard of the reported mid-eclipse epoch (i.e., TDB or UTC).} $M_{2,\rm min}$ is the minimum donor mass from the orbital period alone (Section~\ref{sec:minmass}). \label{tab:known}}
\setlength{\tabcolsep}{4pt}
\begin{tabular}{lccccccccc}
\hline\hline
Name & RA & Dec & $P_{\rm orb}$ & $T_0$ & $t_{\rm ecl}$ & $M_{2,\rm min}$ & $g$ & $r$ & Ref. \\
     & (deg) & (deg) & (min) & (MBJD$_{\rm TDB}$) & (min) & ($M_\odot$) & (mag) & (mag) &  \\
\hline
ZTF J0729$-$0602  & 112.282 & $-$6.046  & 25.92 & \nodata & \nodata & \rev{0.022} & 20.2 & 19.9 & 1 \\
YZ LMi            & 141.661 & $+$36.401 & \rev{28.312233476(10)} & \rev{55469.3164759(6)} & \nodata & \rev{0.020} & 19.2 & 19.5 & 2\rev{, 7} \\
ZTF J0701+5023    & 105.316 & $+$50.389 & 32.47 & \nodata & \nodata & \rev{0.017} & 20.4 & 20.6 & 1, 3 \\
ZTF J0407$-$0007  & 61.955  & $-$0.121  & \rev{35.406897(4)} & \rev{58386.49699(2)} & \nodata & \rev{0.015} & 19.4 & 19.5 & 4 \\
ZTF J2252$-$0519  & 343.154 & $-$5.321  & \rev{37.394112(1)} & \rev{58469.14845(7)} & \nodata & \rev{0.014} & 19.0 & 19.2 & 4 \\
ZTF J1838+0744    & 279.696 & $+$7.746  & 39.61 & \nodata & \nodata & \rev{0.013} & 20.4 & 20.4 & 1 \\
Gaia14aae          & 242.892 & $+$63.142 & \rev{49.70818241(1)} & \rev{58073.8427825(3)} & \nodata & \rev{0.010} & 18.5 & 18.8 & 5\rev{, 7} \\
ZTF J0220+2141    & 35.036  & $+$21.699 & \rev{53.339678(3)} & \rev{58428.24931(8)} & \nodata & \rev{0.009} & 19.7 & 19.8 & 4 \\
SRGe J0453+6224   & 73.501  & $+$62.413 & \rev{55.0802(3)} & \rev{59967.226794129(1)} & \nodata & \rev{0.008} & 18.7 & 18.8 & 6 \\
ZTF J0003+1404    & 0.843   & $+$14.083 & \rev{55.5328(1)} & \rev{58322.52453(1)} & \nodata & \rev{0.008} & 20.2 & 20.2 & 4 \\
ZTF J1637+4917    & 249.431 & $+$49.295 & \rev{61.49919(1)} & \rev{58370.23498(4)} & \nodata & \rev{0.007} & 19.3 & 19.5 & 4 \\
\hline
ZTF\,J1302$-$1746 & 195.532 & $-$17.778 & 39.275554(5) & 61184.13142(1) & 1.95(5) & 0.013 & 19.6 & 19.8 & This work\rev{, 8} \\
ZTF\,J1402+4812   & 210.707 & $+$48.201 & 44.919029(8) & 58768.04395(8) & 1.4 & 0.011 & 21.3 & 21.5 & This work \\
ZTF\,J0106+4328   & 16.723  & $+$43.467 & 45.023040(7) & 59207.12794(7) & 1.8 & 0.011 & 20.8 & 21.0 & This work \\
ZTF\,J2211+2958   & 332.808 & $+$29.980 & 45.336600(4) & 59522.72492(7) & 1.6 & 0.011 & 21.0 & 21.2 & This work \\
ZTF\,J1958+5419   & 299.509 & $+$54.327 & 57.932885(12) & 59119.09385(21) & 2.8 & 0.008 & 21.3 & 21.3 & This work \\
ZTF\,J0859$-$1110 & 134.893 & $-$11.182 & 59.909674(9) & 60050.11478(9) & 0.7 & 0.007 & 19.4 & 19.4 & This work\rev{, 8} \\
\hline
\end{tabular}
\begin{minipage}{0.95\textwidth}
\vspace{4pt}
\footnotesize
References: (1) \citet{2024AandA...683L..10K}; (2) \citet{2005AJ....130.2230A};
(3) \citet{2021AJ....162..113V}; (4) \citet{2022MNRAS.512.5440V};
(5) \citet{2015MNRAS.452.1060C}; (6) \citet{2023ApJ...954...63R}\rev{; (7) \citet{2026MNRAS.548ag673G}; (8) van Roestel et al. (in prep)}.
\end{minipage}
\end{table*}

\section{Conclusion}\label{sec:conclusion}

In this \rev{paper}, we have presented six new eclipsing AM CVn binaries \rev{discovered} in a systematic GPU-accelerated BLS search of ZTF photometry (Householder et al., in prep). Because our search covered a large sample of Pan-STARRS sources rather than a catalog of known white dwarfs, it recovered systems that lack Gaia parallaxes or are absent from published white dwarf catalogs, and we obtained Keck/LRIS spectroscopy of all six systems. \rev{The search found additional candidates with periods below 70 minutes that await spectroscopic follow-up, indicating that many more eclipsing AM CVn binaries should still be detectable in ZTF.} Our main findings are as follows:

\begin{itemize}
\item The six new systems have orbital periods of 39.3 to 59.9 minutes, populating the long-period end of the AM CVn period distribution. They increase the number of AM CVn in which the accreting white dwarf is both the dominant source of optical light and eclipsed by the donor from 11 to 17. 

\item The phase-folded ZTF light curves show eclipses on flat out-of-eclipse baselines. \rev{Their short duration is} consistent with the donor eclipsing the accreting white dwarf, which dominates the optical light at these long orbital periods.

\item The LRIS spectra show prominent He\,\textsc{i} emission with no detectable hydrogen, confirming the AM CVn classification. Several systems show double-peaked helium emission characteristic of an accretion disk. ZTF\,J0859$-$1110, the longest-period system in our sample, shows predominantly single-peaked helium emission that may be analogous to the central spike seen in GP\,Com and V396\,Hya.
\end{itemize}

These discoveries show that wide-field photometric surveys with high cadence can find new eclipsing ultracompact binaries that are missed by parallax-based or catalog-based selection methods. Follow-up high-speed photometry and phase-resolved spectroscopy of these systems are underway and will be presented in a forthcoming paper. With high-cadence light curves we will model the eclipse profiles to constrain the component masses and radii, test whether the donor properties are consistent with a white dwarf, a semi-degenerate helium star, or an evolved cataclysmic variable donor, and place new constraints on the formation channels of AM CVn binaries at long orbital periods. High-cadence eclipse timing observations will also constrain orbital period derivatives and mass-transfer rates, which provide an independent test of the evolutionary state of these systems.

\begin{acknowledgments}
\rev{We thank the anonymous referee for their helpful comments and suggestions, which significantly improved the quality of this work. AH thanks Sunny Wong for a useful conversation during a visit to MIT.} This paper includes data gathered with the 6.5 meter Magellan Telescopes located at Las Campanas Observatory, Chile. This work was supported by a NASA Keck PI Data Award, administered by the NASA Exoplanet Science Institute. Data presented herein were obtained at the W. M. Keck Observatory from telescope time allocated to the National Aeronautics and Space Administration through the agency's scientific partnership with the California Institute of Technology and the University of California. The Observatory was made possible by the generous financial support of the W. M. Keck Foundation. JGM gratefully acknowledges support from the Heising-Simons Foundation and the Pappalardo family through the MIT Pappalardo Fellowship in Physics.

This work is based on observations obtained with the Samuel Oschin Telescope 48-inch and the 60-inch Telescope at the Palomar Observatory as part of the Zwicky Transient Facility project. ZTF is supported by the National Science Foundation under Grants No. AST-1440341 and AST-2034437 and a collaboration including current partners Caltech, IPAC, the Weizmann Institute of Science, the Oskar Klein Center at Stockholm University, the University of Maryland, Deutsches Elektronen-Synchrotron and Humboldt University, the TANGO Consortium of Taiwan, the University of Wisconsin at Milwaukee, Trinity College Dublin, Lawrence Livermore National Laboratories, IN2P3, University of Warwick, Ruhr University Bochum, Northwestern University and former partners the University of Washington, Los Alamos National Laboratories, and Lawrence Berkeley National Laboratories. Operations are conducted by COO, IPAC, and UW.

The authors wish to recognize and acknowledge the very significant cultural role and reverence that the summit of Maunakea has always had within the Native Hawaiian community. We are most fortunate to have the opportunity to conduct observations from this mountain.

Claude Opus and Claude Fable (Anthropic) were used for coding assistance and for help with editing the text at the sentence level. All scientific content, analysis, and conclusions are the authors' own, and the authors take responsibility for all text.

\end{acknowledgments}

\facilities{PO:1.2m (ZTF), Keck:I (LRIS), Magellan:Clay (proto-Lightspeed)}
\software{astropy \citep{2022ApJ...935..167A}, numpy \citep{2020Natur.585..357H}, scipy \citep{2020NaMet..17..261V}, matplotlib \citep{2007CSE.....9...90H}, BLS \citep{2002AandA...391..369K}, ellc \citep{2016AandA...591A.111M}, LPipe \citep{2019PASP..131h4503P}}

\appendix\label{app:helium}
\rev{\section{Helium Line Profiles}}

\onecolumngrid
\begin{center}
    \includegraphics[width=\linewidth]{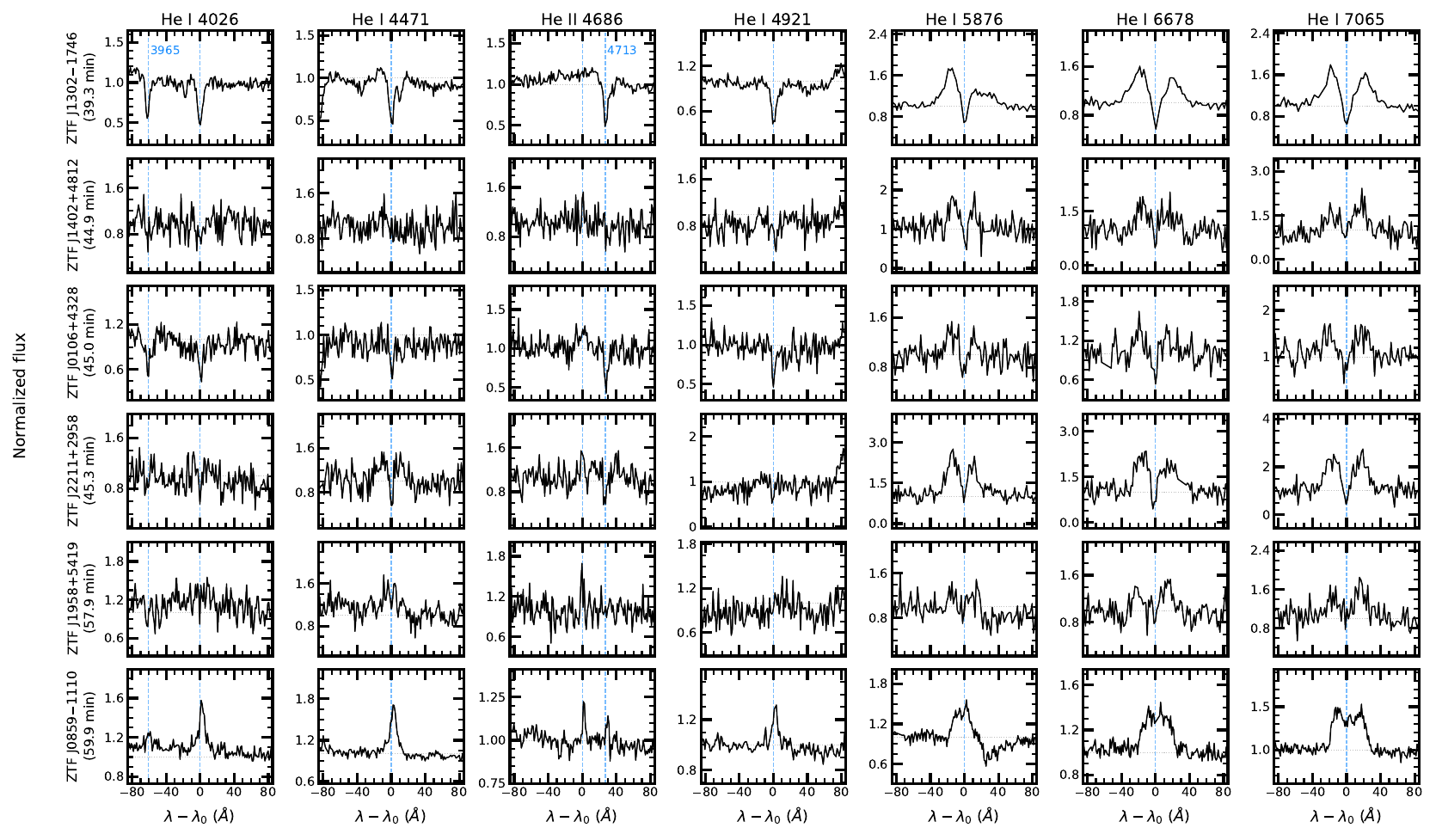}
    \captionof{figure}{\rev{A zoom in to the Keck/LRIS spectra of seven helium lines in the six new systems.}}
    \label{fig:hezoom}
\end{center}
\twocolumngrid

\bibliography{references}{}

\begin{thebibliography}{}
\expandafter\ifx\csname natexlab\endcsname\relax\def\natexlab#1{#1}\fi
\providecommand{\url}[1]{\href{#1}{#1}}
\providecommand{\dodoi}[1]{doi:~\href{http://doi.org/#1}{\nolinkurl{#1}}}
\providecommand{\doeprint}[1]{\href{http://ascl.net/#1}{\nolinkurl{http://ascl.net/#1}}}
\providecommand{\doarXiv}[1]{\href{https://arxiv.org/abs/#1}{\nolinkurl{https://arxiv.org/abs/#1}}}

\bibitem[{S.~F. {Anderson} {et~al.}(2005){Anderson}, {Haggard}, {Homer},
  {Joshi}, {Margon}, {Silvestri}, {Szkody}, {Wolfe}, {Agol}, {Becker},
  {Henden}, {Hall}, {Knapp}, {Richmond}, {Schneider}, {Stinson}, {Barentine},
  {Brewington}, {Brinkmann}, {Harvanek}, {Kleinman}, {Krzesinski}, {Long},
  {Neilsen}, {Nitta}, \& {Snedden}}]{2005AJ....130.2230A}
{Anderson}, S.~F., {Haggard}, D., {Homer}, L., {et~al.} 2005,
  \bibinfo{title}{{Ultracompact AM Canum Venaticorum Binaries from the Sloan
  Digital Sky Survey: Three Candidates Plus the First Confirmed Eclipsing
  System},} \aj, 130, 2230, \dodoi{10.1086/491587}

\bibitem[{ {Astropy Collaboration} {et~al.}(2022){Astropy Collaboration},
  {Price-Whelan}, {Lim}, {Earl}, {Starkman}, {Bradley}, {Shupe}, {Patil},
  {Corrales}, {Brasseur}, {N{\"o}the}, {Donath}, {Tollerud}, {Morris},
  {Ginsburg}, {Vaher}, {Weaver}, {Tocknell}, {Jamieson}, {van Kerkwijk},
  {Robitaille}, {Merry}, {Bachetti}, {G{\"u}nther}, {Aldcroft},
  {Alvarado-Montes}, {Archibald}, {B{\'o}di}, {Bapat}, {Barentsen},
  {Baz{\'a}n}, {Biswas}, {Boquien}, {Burke}, {Cara}, {Cara}, {Conroy},
  {Conseil}, {Craig}, {Cross}, {Cruz}, {D'Eugenio}, {Dencheva}, {Devillepoix},
  {Dietrich}, {Eigenbrot}, {Erben}, {Ferreira}, {Foreman-Mackey}, {Fox},
  {Freij}, {Garg}, {Geda}, {Glattly}, {Gondhalekar}, {Gordon}, {Grant},
  {Greenfield}, {Groener}, {Guest}, {Gurovich}, {Handberg}, {Hart},
  {Hatfield-Dodds}, {Homeier}, {Hosseinzadeh}, {Jenness}, {Jones}, {Joseph},
  {Kalmbach}, {Karamehmetoglu}, {Ka{\l}uszy{\'n}ski}, {Kelley}, {Kern},
  {Kerzendorf}, {Koch}, {Kulumani}, {Lee}, {Ly}, {Ma}, {MacBride}, {Maljaars},
  {Muna}, {Murphy}, {Norman}, {O'Steen}, {Oman}, {Pacifici}, {Pascual},
  {Pascual-Granado}, {Patil}, {Perren}, {Pickering}, {Rastogi}, {Roulston},
  {Ryan}, {Rykoff}, {Sabater}, {Sakurikar}, {Salgado}, {Sanghi}, {Saunders},
  {Savchenko}, {Schwardt}, {Seifert-Eckert}, {Shih}, {Jain}, {Shukla}, {Sick},
  {Simpson}, {Singanamalla}, {Singer}, {Singhal}, {Sinha}, {Sip{\H{o}}cz},
  {Spitler}, {Stansby}, {Streicher}, {{\v{S}}umak}, {Swinbank}, {Taranu},
  {Tewary}, {Tremblay}, {de Val-Borro}, {Van Kooten}, {Vasovi{\'c}}, {Verma},
  {de Miranda Cardoso}, {Williams}, {Wilson}, {Winkel}, {Wood-Vasey}, {Xue},
  {Yoachim}, {Zhang}, {Zonca}, \& {Astropy Project
  Contributors}}]{2022ApJ...935..167A}
{Astropy Collaboration}, {Price-Whelan}, A.~M., {Lim}, P.~L., {et~al.} 2022,
  \bibinfo{title}{{The Astropy Project: Sustaining and Growing a
  Community-oriented Open-source Project and the Latest Major Release (v5.0) of
  the Core Package},} \apj, 935, 167, \dodoi{10.3847/1538-4357/ac7c74}

\bibitem[{E.~C. {Bellm} {et~al.}(2019){Bellm}, {Kulkarni}, {Graham}, {Dekany},
  {Smith}, {Riddle}, {Masci}, {Helou}, {Prince}, {Adams}, {Barbarino},
  {Barlow}, {Bauer}, {Beck}, {Belicki}, {Biswas}, {Blagorodnova}, {Bodewits},
  {Bolin}, {Brinnel}, {Brooke}, {Bue}, {Bulla}, {Burruss}, {Cenko}, {Chang},
  {Connolly}, {Coughlin}, {Cromer}, {Cunningham}, {De}, {Delacroix}, {Desai},
  {Duev}, {Eadie}, {Farnham}, {Feeney}, {Feindt}, {Flynn}, {Franckowiak},
  {Frederick}, {Fremling}, {Gal-Yam}, {Gezari}, {Giomi}, {Goldstein},
  {Golkhou}, {Goobar}, {Groom}, {Hacopians}, {Hale}, {Henning}, {Ho}, {Hover},
  {Howell}, {Hung}, {Huppenkothen}, {Imel}, {Ip}, {Ivezi{\'c}}, {Jackson},
  {Jones}, {Juric}, {Kasliwal}, {Kaspi}, {Kaye}, {Kelley}, {Kowalski},
  {Kramer}, {Kupfer}, {Landry}, {Laher}, {Lee}, {Lin}, {Lin}, {Lunnan},
  {Giomi}, {Mahabal}, {Mao}, {Miller}, {Monkewitz}, {Murphy}, {Ngeow},
  {Nordin}, {Nugent}, {Ofek}, {Patterson}, {Penprase}, {Porter}, {Rauch},
  {Rebbapragada}, {Reiley}, {Rigault}, {Rodriguez}, {van Roestel}, {Rusholme},
  {van Santen}, {Schulze}, {Shupe}, {Singer}, {Soumagnac}, {Stein}, {Surace},
  {Sollerman}, {Szkody}, {Taddia}, {Terek}, {Van Sistine}, {van Velzen},
  {Vestrand}, {Walters}, {Ward}, {Ye}, {Yu}, {Yan}, \&
  {Zolkower}}]{2019PASP..131a8002B}
{Bellm}, E.~C., {Kulkarni}, S.~R., {Graham}, M.~J., {et~al.} 2019,
  \bibinfo{title}{{The Zwicky Transient Facility: System Overview, Performance,
  and First Results},} \pasp, 131, 018002, \dodoi{10.1088/1538-3873/aaecbe}

\bibitem[{L. {Bildsten} {et~al.}(2006){Bildsten}, {Townsley}, {Deloye}, \&
  {Nelemans}}]{2006ApJ...640..466B}
{Bildsten}, L., {Townsley}, D.~M., {Deloye}, C.~J., \& {Nelemans}, G. 2006,
  \bibinfo{title}{{The Thermal State of the Accreting White Dwarf in AM Canum
  Venaticorum Binaries},} \apj, 640, 466, \dodoi{10.1086/500080}

\bibitem[{K. {B{\k{a}}kowska} {et~al.}(2021){B{\k{a}}kowska}, {Marsh},
  {Steeghs}, {Nelemans}, \& {Groot}}]{2021AandA...645A.114B}
{B{\k{a}}kowska}, K., {Marsh}, T.~R., {Steeghs}, D., {Nelemans}, G., \&
  {Groot}, P.~J. 2021, \bibinfo{title}{{Spectroscopy of the helium-rich binary
  ES Ceti reveals accretion via a disc and evidence of eclipses},} \aap, 645,
  A114, \dodoi{10.1051/0004-6361/202039266}

\bibitem[{H.~C. {Campbell} {et~al.}(2015){Campbell}, {Marsh}, {Fraser},
  {Hodgkin}, {de Miguel}, {G{\"a}nsicke}, {Steeghs}, {Hourihane}, {Breedt},
  {Littlefair}, {Koposov}, {Wyrzykowski}, {Altavilla}, {Blagorodnova},
  {Clementini}, {Damljanovic}, {Delgado}, {Dennefeld}, {Drake},
  {Fern{\'a}ndez-Hern{\'a}ndez}, {Gilmore}, {Gualandi}, {Hamanowicz},
  {Handzlik}, {Hardy}, {Harrison}, {I{\l}kiewicz}, {Jonker}, {Kochanek},
  {Ko{\l}aczkowski}, {Kostrzewa-Rutkowska}, {Kotak}, {van Leeuwen}, {Leto},
  {Ochner}, {Pawlak}, {Palaversa}, {Rixon}, {Rybicki}, {Shappee}, {Smartt},
  {Torres}, {Tomasella}, {Turatto}, {Ulaczyk}, {van Velzen}, {Vince}, {Walton},
  {Wielg{\'o}rski}, {Wevers}, {Whitelock}, {Yoldas}, {De Angeli}, {Burgess},
  {Busso}, {Busuttil}, {Butterley}, {Chambers}, {Copperwheat}, {Danilet},
  {Dhillon}, {Evans}, {Eyer}, {Froebrich}, {Gomboc}, {Holland}, {Holoien},
  {Jarvis}, {Kaiser}, {Kann}, {Koester}, {Kolb}, {Komossa}, {Magnier},
  {Mahabal}, {Polshaw}, {Prieto}, {Prusti}, {Riello}, {Scholz}, {Simonian},
  {Stanek}, {Szabados}, {Waters}, \& {Wilson}}]{2015MNRAS.452.1060C}
{Campbell}, H.~C., {Marsh}, T.~R., {Fraser}, M., {et~al.} 2015,
  \bibinfo{title}{{Total eclipse of the heart: the AM CVn
  Gaia14aae/ASSASN-14cn},} \mnras, 452, 1060, \dodoi{10.1093/mnras/stv1224}

\bibitem[{P.~J. {Carter} {et~al.}(2013){Carter}, {Marsh}, {Steeghs}, {Groot},
  {Nelemans}, {Levitan}, {Rau}, {Copperwheat}, {Kupfer}, \&
  {Roelofs}}]{2013MNRAS.429.2143C}
{Carter}, P.~J., {Marsh}, T.~R., {Steeghs}, D., {et~al.} 2013,
  \bibinfo{title}{{A search for the hidden population of AM CVn binaries in the
  Sloan Digital Sky Survey},} \mnras, 429, 2143, \dodoi{10.1093/mnras/sts485}

\bibitem[{J. {Chakraborty} {et~al.}(2024){Chakraborty}, {Burdge}, {Rappaport},
  {Munday}, {Chen}, {Rodr{\'i}guez-Gil}, {Dhillon}, {Hughes}, {Nelemans},
  {Kara}, {Bellm}, {Brown}, {Castro Segura}, {Chen}, {Chickles}, {Dyer},
  {Dekany}, {Drake}, {Garbutt}, {Graham}, {Green}, {Jarvis}, {Kennedy},
  {Kerry}, {Kulkarni}, {Littlefair}, {Mahabal}, {Masci}, {McCormac}, {Parsons},
  {Pelisoli}, {Pike}, {Prince}, {Riddle}, {van Roestel}, {Sahman}, {Wold}, \&
  {Wong}}]{2024ApJ...977..262C}
{Chakraborty}, J., {Burdge}, K.~B., {Rappaport}, S.~A., {et~al.} 2024,
  \bibinfo{title}{{Expanding the Ultracompacts: Gravitational-wave-driven Mass
  Transfer in the Shortest-period Binaries with Accretion Disks},} \apj, 977,
  262, \dodoi{10.3847/1538-4357/ad9563}

\bibitem[{K.~C. {Chambers} {et~al.}(2016){Chambers}, {Magnier}, {Metcalfe},
  {Flewelling}, {Huber}, {Waters}, {Denneau}, {Draper}, {Farrow}, {Finkbeiner},
  {Holmberg}, {Koppenhoefer}, {Price}, {Rest}, {Saglia}, {Schlafly}, {Smartt},
  {Sweeney}, {Wainscoat}, {Burgett}, {Chastel}, {Grav}, {Heasley}, {Hodapp},
  {Jedicke}, {Kaiser}, {Kudritzki}, {Luppino}, {Lupton}, {Monet}, {Morgan},
  {Onaka}, {Shiao}, {Stubbs}, {Tonry}, {White}, {Ba{\~n}ados}, {Bell},
  {Bender}, {Bernard}, {Boegner}, {Boffi}, {Botticella}, {Calamida},
  {Casertano}, {Chen}, {Chen}, {Cole}, {Deacon}, {Frenk}, {Fitzsimmons},
  {Gezari}, {Gibbs}, {Goessl}, {Goggia}, {Gourgue}, {Goldman}, {Grant},
  {Grebel}, {Hambly}, {Hasinger}, {Heavens}, {Heckman}, {Henderson}, {Henning},
  {Holman}, {Hopp}, {Ip}, {Isani}, {Jackson}, {Keyes}, {Koekemoer}, {Kotak},
  {Le}, {Liska}, {Long}, {Lucey}, {Liu}, {Martin}, {Masci}, {McLean}, {Mindel},
  {Misra}, {Morganson}, {Murphy}, {Obaika}, {Narayan}, {Nieto-Santisteban},
  {Norberg}, {Peacock}, {Pier}, {Postman}, {Primak}, {Rae}, {Rai}, {Riess},
  {Riffeser}, {Rix}, {R{\"o}ser}, {Russel}, {Rutz}, {Schilbach}, {Schultz},
  {Scolnic}, {Strolger}, {Szalay}, {Seitz}, {Small}, {Smith}, {Soderblom},
  {Taylor}, {Thomson}, {Taylor}, {Thakar}, {Thiel}, {Thilker}, {Unger},
  {Urata}, {Valenti}, {Wagner}, {Walder}, {Walter}, {Watters}, {Werner},
  {Wood-Vasey}, \& {Wyse}}]{2016arXiv161205560C}
{Chambers}, K.~C., {Magnier}, E.~A., {Metcalfe}, N., {et~al.} 2016,
  \bibinfo{title}{{The Pan-STARRS1 Surveys},} arXiv e-prints, arXiv:1612.05560,
  \dodoi{10.48550/arXiv.1612.05560}

\bibitem[{E.~T. {Chickles} {et~al.}(2026){Chickles}, {Chakraborty}, {Burdge},
  {Dhillon}, {Draghis}, {El-Badry}, {Green}, {Householder}, {Hughes}, {Layden},
  {Littlefair}, {Munday}, {Pelisoli}, {Redden}, {Tonry}, {van Roestel},
  {Angile}, {Brown}, {Castro Segura}, {Dinsmore}, {Dyer}, {Furesz}, {Gabutti},
  {Garbutt}, {Garc{\'\i}a-Mej{\'\i}a}, {Jarvis}, {Kennedy}, {Kerry},
  {McCormac}, {Mo}, {Osip}, {Parsons}, {Pike}, {Piotrowski}, {Romani},
  {Sahman}, \& {Simcoe}}]{2026ApJ..1000..237C}
{Chickles}, E.~T., {Chakraborty}, J., {Burdge}, K.~B., {et~al.} 2026,
  \bibinfo{title}{{An Eclipsing 8.56 Minutes Orbital Period Mass-transferring
  Binary},} \apj, 1000, 237, \dodoi{10.3847/1538-4357/ae4871}

\bibitem[{M.~C. {Cook} \& B. {Warner}(1984){Cook} \&
  {Warner}}]{1984MNRAS.207..705C}
{Cook}, M.~C., \& {Warner}, B. 1984, \bibinfo{title}{{High-speed photometry of
  Z Chamaeleontis in the quiescent state.},} \mnras, 207, 705,
  \dodoi{10.1093/mnras/207.4.705}

\bibitem[{C.~M. {Copperwheat} {et~al.}(2011){Copperwheat}, {Marsh},
  {Littlefair}, {Dhillon}, {Ramsay}, {Drake}, {G{\"a}nsicke}, {Groot},
  {Hakala}, {Koester}, {Nelemans}, {Roelofs}, {Southworth}, {Steeghs}, \&
  {Tulloch}}]{2011MNRAS.410.1113C}
{Copperwheat}, C.~M., {Marsh}, T.~R., {Littlefair}, S.~P., {et~al.} 2011,
  \bibinfo{title}{{SDSS J0926+3624: the shortest period eclipsing binary
  star},} \mnras, 410, 1113, \dodoi{10.1111/j.1365-2966.2010.17508.x}

\bibitem[{C.~J. {Deloye} {et~al.}(2007){Deloye}, {Taam}, {Winisdoerffer}, \&
  {Chabrier}}]{2007MNRAS.381..525D}
{Deloye}, C.~J., {Taam}, R.~E., {Winisdoerffer}, C., \& {Chabrier}, G. 2007,
  \bibinfo{title}{{The thermal evolution of the donors in AM Canum Venaticorum
  binaries},} \mnras, 381, 525, \dodoi{10.1111/j.1365-2966.2007.12262.x}

\bibitem[{P.~P. {Eggleton}(1983){Eggleton}}]{1983ApJ...268..368E}
{Eggleton}, P.~P. 1983, \bibinfo{title}{{Aproximations to the radii of Roche
  lobes.},} \apj, 268, 368, \dodoi{10.1086/160960}

\bibitem[{H.~A. {Flewelling} {et~al.}(2020){Flewelling}, {Magnier}, {Chambers},
  {Heasley}, {Holmberg}, {Huber}, {Sweeney}, {Waters}, {Calamida}, {Casertano},
  {Chen}, {Farrow}, {Hasinger}, {Henderson}, {Long}, {Metcalfe}, {Narayan},
  {Nieto-Santisteban}, {Norberg}, {Rest}, {Saglia}, {Szalay}, {Thakar},
  {Tonry}, {Valenti}, {Werner}, {White}, {Denneau}, {Draper}, {Hodapp},
  {Jedicke}, {Kaiser}, {Kudritzki}, {Price}, {Wainscoat}, {Chastel}, {McLean},
  {Postman}, \& {Shiao}}]{2020ApJS..251....7F}
{Flewelling}, H.~A., {Magnier}, E.~A., {Chambers}, K.~C., {et~al.} 2020,
  \bibinfo{title}{{The Pan-STARRS1 Database and Data Products},} \apjs, 251, 7,
  \dodoi{10.3847/1538-4365/abb82d}

\bibitem[{N.~P. {Gentile Fusillo} {et~al.}(2021){Gentile Fusillo}, {Tremblay},
  {Cukanovaite}, {Vorontseva}, {Lallement}, {Hollands}, {G{\"a}nsicke},
  {Burdge}, {McCleery}, \& {Jordan}}]{2021MNRAS.508.3877G}
{Gentile Fusillo}, N.~P., {Tremblay}, P.-E., {Cukanovaite}, E., {et~al.} 2021,
  \bibinfo{title}{{A catalogue of white dwarfs in Gaia EDR3},} \mnras, 508,
  3877, \dodoi{10.1093/mnras/stab2672}

\bibitem[{M.~J. {Graham} {et~al.}(2019){Graham}, {Kulkarni}, {Bellm}, {Adams},
  {Barbarino}, {Blagorodnova}, {Bodewits}, {Bolin}, {Brady}, {Cenko}, {Chang},
  {Coughlin}, {De}, {Eadie}, {Farnham}, {Feindt}, {Franckowiak}, {Fremling},
  {Gezari}, {Ghosh}, {Goldstein}, {Golkhou}, {Goobar}, {Ho}, {Huppenkothen},
  {Ivezi{\'c}}, {Jones}, {Juric}, {Kaplan}, {Kasliwal}, {Kelley}, {Kupfer},
  {Lee}, {Lin}, {Lunnan}, {Mahabal}, {Miller}, {Ngeow}, {Nugent}, {Ofek},
  {Prince}, {Rauch}, {van Roestel}, {Schulze}, {Singer}, {Sollerman}, {Taddia},
  {Yan}, {Ye}, {Yu}, {Barlow}, {Bauer}, {Beck}, {Belicki}, {Biswas}, {Brinnel},
  {Brooke}, {Bue}, {Bulla}, {Burruss}, {Connolly}, {Cromer}, {Cunningham},
  {Dekany}, {Delacroix}, {Desai}, {Duev}, {Feeney}, {Flynn}, {Frederick},
  {Gal-Yam}, {Giomi}, {Groom}, {Hacopians}, {Hale}, {Helou}, {Henning},
  {Hover}, {Hillenbrand}, {Howell}, {Hung}, {Imel}, {Ip}, {Jackson}, {Kaspi},
  {Kaye}, {Kowalski}, {Kramer}, {Kuhn}, {Landry}, {Laher}, {Mao}, {Masci},
  {Monkewitz}, {Murphy}, {Nordin}, {Patterson}, {Penprase}, {Porter},
  {Rebbapragada}, {Reiley}, {Riddle}, {Rigault}, {Rodriguez}, {Rusholme}, {van
  Santen}, {Shupe}, {Smith}, {Soumagnac}, {Stein}, {Surace}, {Szkody}, {Terek},
  {Van Sistine}, {van Velzen}, {Vestrand}, {Walters}, {Ward}, {Zhang}, \&
  {Zolkower}}]{2019PASP..131g8001G}
{Graham}, M.~J., {Kulkarni}, S.~R., {Bellm}, E.~C., {et~al.} 2019,
  \bibinfo{title}{{The Zwicky Transient Facility: Science Objectives},} \pasp,
  131, 078001, \dodoi{10.1088/1538-3873/ab006c}

\bibitem[{M.~J. {Green} {et~al.}(2025){Green}, {van Roestel}, \&
  {Wong}}]{2025AandA...700A.107G}
{Green}, M.~J., {van Roestel}, J., \& {Wong}, T. L.~S. 2025, \bibinfo{title}{{A
  catalogue of ultracompact mass-transferring white dwarf binaries},} \aap,
  700, A107, \dodoi{10.1051/0004-6361/202554925}

\bibitem[{M.~J. {Green} {et~al.}(2018){Green}, {Marsh}, {Steeghs}, {Kupfer},
  {Ashley}, {Bloemen}, {Breedt}, {Campbell}, {Chakpor}, {Copperwheat},
  {Dhillon}, {Hallinan}, {Hardy}, {Hermes}, {Kerry}, {Littlefair}, {Milburn},
  {Parsons}, {Prasert}, {van Roestel}, {Sahman}, \&
  {Singh}}]{2018MNRAS.476.1663G}
{Green}, M.~J., {Marsh}, T.~R., {Steeghs}, D.~T.~H., {et~al.} 2018,
  \bibinfo{title}{{High-speed photometry of Gaia14aae: an eclipsing AM CVn that
  challenges formation models},} \mnras, 476, 1663,
  \dodoi{10.1093/mnras/sty299}

\bibitem[{M.~J. {Green} {et~al.}(2026){Green}, {Marsh}, {van Roestel}, {Wong},
  {Belloni}, {Kilic}, {Breedt}, {Brown}, {Copperwheat}, {Chakpor}, {Dhillon},
  {Segura}, {Dyer}, {Garbutt}, {Jarvis}, {Kengkriangkrai}, {Kennedy}, {Kerry},
  {Kupfer}, {Littlefair}, {McCormac}, {Munday}, {Parsons}, {Pike}, {Pelisoli},
  {Rodr{\'\i}guez-Gil}, {Sahman}, \& {Yates}}]{2026MNRAS.548ag673G}
{Green}, M.~J., {Marsh}, T.~R., {van Roestel}, J., {et~al.} 2026,
  \bibinfo{title}{{No period change in two long-period AM CVn binaries},}
  \mnras, 548, stag673, \dodoi{10.1093/mnras/stag673}

\bibitem[{C.~R. {Harris} {et~al.}(2020){Harris}, {Millman}, {van der Walt},
  {Gommers}, {Virtanen}, {Cournapeau}, {Wieser}, {Taylor}, {Berg}, {Smith},
  {Kern}, {Picus}, {Hoyer}, {van Kerkwijk}, {Brett}, {Haldane}, {del R{\'\i}o},
  {Wiebe}, {Peterson}, {G{\'e}rard-Marchant}, {Sheppard}, {Reddy}, {Weckesser},
  {Abbasi}, {Gohlke}, \& {Oliphant}}]{2020Natur.585..357H}
{Harris}, C.~R., {Millman}, K.~J., {van der Walt}, S.~J., {et~al.} 2020,
  \bibinfo{title}{{Array programming with NumPy},} \nat, 585, 357,
  \dodoi{10.1038/s41586-020-2649-2}

\bibitem[{K. {Horne} \& T.~R. {Marsh}(1986){Horne} \&
  {Marsh}}]{1986MNRAS.218..761H}
{Horne}, K., \& {Marsh}, T.~R. 1986, \bibinfo{title}{{Emission line formation
  in accretion discs},} \mnras, 218, 761, \dodoi{10.1093/mnras/218.4.761}

\bibitem[{J.~D. {Hunter}(2007){Hunter}}]{2007CSE.....9...90H}
{Hunter}, J.~D. 2007, \bibinfo{title}{{Matplotlib: A 2D Graphics Environment},}
  Computing in Science and Engineering, 9, 90, \dodoi{10.1109/MCSE.2007.55}

\bibitem[{J.~M. {Khalil} {et~al.}(2024){Khalil}, {van Roestel}, {Bellm},
  {Bloom}, {Dekany}, {Drake}, {Graham}, {Groom}, {Kulkarni}, {Laher},
  {Mahabal}, {Prince}, \& {Riddle}}]{2024AandA...683L..10K}
{Khalil}, J.~M., {van Roestel}, J., {Bellm}, E.~C., {et~al.} 2024,
  \bibinfo{title}{{Four new eclipsing accreting ultracompact white dwarf
  binaries found with the Zwicky Transient Facility},} \aap, 683, L10,
  \dodoi{10.1051/0004-6361/202349010}

\bibitem[{G. {Kov{\'a}cs} {et~al.}(2002){Kov{\'a}cs}, {Zucker}, \&
  {Mazeh}}]{2002AandA...391..369K}
{Kov{\'a}cs}, G., {Zucker}, S., \& {Mazeh}, T. 2002, \bibinfo{title}{{A
  box-fitting algorithm in the search for periodic transits},} \aap, 391, 369,
  \dodoi{10.1051/0004-6361:20020802}

\bibitem[{T. {Kupfer} {et~al.}(2013){Kupfer}, {Groot}, {Levitan}, {Steeghs},
  {Marsh}, {Rutten}, \& {Nelemans}}]{2013MNRAS.432.2048K}
{Kupfer}, T., {Groot}, P.~J., {Levitan}, D., {et~al.} 2013,
  \bibinfo{title}{{Orbital periods and accretion disc structure of four AM CVn
  systems},} \mnras, 432, 2048, \dodoi{10.1093/mnras/stt524}

\bibitem[{T. {Kupfer} {et~al.}(2016){Kupfer}, {Steeghs}, {Groot}, {Marsh},
  {Nelemans}, \& {Roelofs}}]{2016MNRAS.457.1828K}
{Kupfer}, T., {Steeghs}, D., {Groot}, P.~J., {et~al.} 2016,
  \bibinfo{title}{{UVES and X-Shooter spectroscopy of the emission line AM CVn
  systems GP Com and V396 Hya},} \mnras, 457, 1828,
  \dodoi{10.1093/mnras/stw126}

\bibitem[{C. {Layden} {et~al.}(2026{\natexlab{a}}){Layden}, {Burdge}, {Furesz},
  {Garc{\'\i}a-Mej{\'\i}a}, {Dinsmore}, {Mo}, {Osip}, {Piotrowski}, {Romani},
  {Berne}, {Chakrabarty}, \& {Chickles}}]{2026PASP..138f5001L}
{Layden}, C., {Burdge}, K., {Furesz}, G., {et~al.} 2026{\natexlab{a}},
  \bibinfo{title}{{proto-Lightspeed: A High-speed, Ultra-low Read Noise Imager
  on the Magellan Clay Telescope},} \pasp, 138, 065001,
  \dodoi{10.1088/1538-3873/ae6dfc}

\bibitem[{C. {Layden} {et~al.}(2026{\natexlab{b}}){Layden}, {Burdge},
  {Piotrowski}, {Furesz}, {Garc{\'\i}a-Mej{\'\i}a}, {Dinsmore}, {Romani},
  {Osip}, {Mo}, {Berger}, {Householder}, {Frostig}, {Chickles}, {Pellecer}, \&
  {Desai}}]{2026SPIE14149E..2IL}
{Layden}, C., {Burdge}, K., {Piotrowski}, J.~J., {et~al.} 2026{\natexlab{b}},
  in Society of Photo-Optical Instrumentation Engineers (SPIE) Conference
  Series, Vol. 14149, Ground-based and Airborne Instrumentation for Astronomy
  XI, ed. J.~D.~R. {Vernet}, J.~J. {Bryant}, \& K.~{Motohara} (SPIE), 141492I,
  \dodoi{10.1117/12.3102886}

\bibitem[{D. {Levitan} {et~al.}(2015){Levitan}, {Groot}, {Prince}, {Kulkarni},
  {Laher}, {Ofek}, {Sesar}, \& {Surace}}]{2015MNRAS.446..391L}
{Levitan}, D., {Groot}, P.~J., {Prince}, T.~A., {et~al.} 2015,
  \bibinfo{title}{{Long-term photometric behaviour of outbursting AM CVn
  systems},} \mnras, 446, 391, \dodoi{10.1093/mnras/stu2105}

\bibitem[{D. {Levitan} {et~al.}(2014){Levitan}, {Kupfer}, {Groot}, {Margon},
  {Prince}, {Kulkarni}, {Hallinan}, {Harding}, {Kyne}, {Laher}, {Ofek},
  {Rutten}, {Sesar}, \& {Surace}}]{2014ApJ...785..114L}
{Levitan}, D., {Kupfer}, T., {Groot}, P.~J., {et~al.} 2014,
  \bibinfo{title}{{PTF1 J191905.19+481506.2{\textemdash}a Partially Eclipsing
  AM CVn System Discovered in the Palomar Transient Factory},} \apj, 785, 114,
  \dodoi{10.1088/0004-637X/785/2/114}

\bibitem[{T.~R. {Marsh}(1999){Marsh}}]{1999MNRAS.304..443M}
{Marsh}, T.~R. 1999, \bibinfo{title}{{Kinematics of the helium accretor GP
  Com},} \mnras, 304, 443, \dodoi{10.1046/j.1365-8711.1999.02323.x}

\bibitem[{F.~J. {Masci} {et~al.}(2019){Masci}, {Laher}, {Rusholme}, {Shupe},
  {Groom}, {Surace}, {Jackson}, {Monkewitz}, {Beck}, {Flynn}, {Terek},
  {Landry}, {Hacopians}, {Desai}, {Howell}, {Brooke}, {Imel}, {Wachter}, {Ye},
  {Lin}, {Cenko}, {Cunningham}, {Rebbapragada}, {Bue}, {Miller}, {Mahabal},
  {Bellm}, {Patterson}, {Juri{\'c}}, {Golkhou}, {Ofek}, {Walters}, {Graham},
  {Kasliwal}, {Dekany}, {Kupfer}, {Burdge}, {Cannella}, {Barlow}, {Van
  Sistine}, {Giomi}, {Fremling}, {Blagorodnova}, {Levitan}, {Riddle}, {Smith},
  {Helou}, {Prince}, \& {Kulkarni}}]{2019PASP..131a8003M}
{Masci}, F.~J., {Laher}, R.~R., {Rusholme}, B., {et~al.} 2019,
  \bibinfo{title}{{The Zwicky Transient Facility: Data Processing, Products,
  and Archive},} \pasp, 131, 018003, \dodoi{10.1088/1538-3873/aae8ac}

\bibitem[{P.~F.~L. {Maxted}(2016){Maxted}}]{2016AandA...591A.111M}
{Maxted}, P.~F.~L. 2016, \bibinfo{title}{{ellc: A fast, flexible light curve
  model for detached eclipsing binary stars and transiting exoplanets},} \aap,
  591, A111, \dodoi{10.1051/0004-6361/201628579}

\bibitem[{G. {Nelemans} {et~al.}(2001){Nelemans}, {Portegies Zwart}, {Verbunt},
  \& {Yungelson}}]{2001AandA...368..939N}
{Nelemans}, G., {Portegies Zwart}, S.~F., {Verbunt}, F., \& {Yungelson}, L.~R.
  2001, \bibinfo{title}{{Population synthesis for double white dwarfs. II.
  Semi-detached systems: AM CVn stars},} \aap, 368, 939,
  \dodoi{10.1051/0004-6361:20010049}

\bibitem[{G. {Nelemans} {et~al.}(2004){Nelemans}, {Yungelson}, \& {Portegies
  Zwart}}]{2004MNRAS.349..181N}
{Nelemans}, G., {Yungelson}, L.~R., \& {Portegies Zwart}, S.~F. 2004,
  \bibinfo{title}{{Short-period AM CVn systems as optical, X-ray and
  gravitational-wave sources},} \mnras, 349, 181,
  \dodoi{10.1111/j.1365-2966.2004.07479.x}

\bibitem[{G. {Nelemans} {et~al.}(2010){Nelemans}, {Yungelson}, {van der Sluys},
  \& {Tout}}]{2010MNRAS.401.1347N}
{Nelemans}, G., {Yungelson}, L.~R., {van der Sluys}, M.~V., \& {Tout}, C.~A.
  2010, \bibinfo{title}{{The chemical composition of donors in AM CVn stars and
  ultracompact X-ray binaries: observational tests of their formation},}
  \mnras, 401, 1347, \dodoi{10.1111/j.1365-2966.2009.15731.x}

\bibitem[{J.~B. {Oke} {et~al.}(1995){Oke}, {Cohen}, {Carr}, {Cromer},
  {Dingizian}, {Harris}, {Labrecque}, {Lucinio}, {Schaal}, {Epps}, \&
  {Miller}}]{1995PASP..107..375O}
{Oke}, J.~B., {Cohen}, J.~G., {Carr}, M., {et~al.} 1995, \bibinfo{title}{{The
  Keck Low-Resolution Imaging Spectrometer},} \pasp, 107, 375,
  \dodoi{10.1086/133562}

\bibitem[{D.~A. {Perley}(2019){Perley}}]{2019PASP..131h4503P}
{Perley}, D.~A. 2019, \bibinfo{title}{{Fully Automated Reduction of Longslit
  Spectroscopy with the Low Resolution Imaging Spectrometer at the Keck
  Observatory},} \pasp, 131, 084503, \dodoi{10.1088/1538-3873/ab215d}

\bibitem[{P. {Podsiadlowski} {et~al.}(2003){Podsiadlowski}, {Han}, \&
  {Rappaport}}]{2003MNRAS.340.1214P}
{Podsiadlowski}, P., {Han}, Z., \& {Rappaport}, S. 2003,
  \bibinfo{title}{{Cataclysmic variables with evolved secondaries and the
  progenitors of AM CVn stars},} \mnras, 340, 1214,
  \dodoi{10.1046/j.1365-8711.2003.06380.x}

\bibitem[{G. {Ramsay} {et~al.}(2018){Ramsay}, {Green}, {Marsh}, {Kupfer},
  {Breedt}, {Korol}, {Groot}, {Knigge}, {Nelemans}, {Steeghs}, {Woudt}, \&
  {Aungwerojwit}}]{2018AandA...620A.141R}
{Ramsay}, G., {Green}, M.~J., {Marsh}, T.~R., {et~al.} 2018,
  \bibinfo{title}{{Physical properties of AM CVn stars: New insights from Gaia
  DR2},} \aap, 620, A141, \dodoi{10.1051/0004-6361/201834261}

\bibitem[{C. {Rockosi} {et~al.}(2010){Rockosi}, {Stover}, {Kibrick},
  {Lockwood}, {Peck}, {Cowley}, {Bolte}, {Adkins}, {Alcott}, {Allen}, {Brown},
  {Cabak}, {Deich}, {Hilyard}, {Kassis}, {Lanclos}, {Lewis}, {Pfister},
  {Phillips}, {Robinson}, {Saylor}, {Thompson}, {Ward}, {Wei}, \&
  {Wright}}]{2010SPIE.7735E..0RR}
{Rockosi}, C., {Stover}, R., {Kibrick}, R., {et~al.} 2010, in Society of
  Photo-Optical Instrumentation Engineers (SPIE) Conference Series, Vol. 7735,
  Ground-based and Airborne Instrumentation for Astronomy III, ed. I.~S.
  {McLean}, S.~K. {Ramsay}, \& H.~{Takami}, 77350R, \dodoi{10.1117/12.856818}

\bibitem[{A.~C. {Rodriguez} {et~al.}(2023){Rodriguez}, {Galiullin}, {Gilfanov},
  {Kulkarni}, {Khamitov}, {Bikmaev}, {van Roestel}, {Yungelson}, {El-Badry},
  {Sunayev}, {Prince}, {Buntov}, {Caiazzo}, {Drake}, {Gorbachev}, {Graham},
  {Gumerov}, {Irtuganov}, {Laher}, {Masci}, {Medvedev}, {Purdum},
  {Sakhibullin}, {Sklyanov}, {Smith}, {Szkody}, \&
  {Vanderbosch}}]{2023ApJ...954...63R}
{Rodriguez}, A.~C., {Galiullin}, I., {Gilfanov}, M., {et~al.} 2023,
  \bibinfo{title}{{SRGeJ045359.9+622444: A 55 Minute Period Eclipsing AM Canum
  Venaticorum Star Discovered from a Joint SRG/eROSITA + ZTF Search},} \apj,
  954, 63, \dodoi{10.3847/1538-4357/ace698}

\bibitem[{G.~H.~A. {Roelofs} {et~al.}(2005){Roelofs}, {Groot}, {Marsh},
  {Steeghs}, {Barros}, \& {Nelemans}}]{2005MNRAS.361..487R}
{Roelofs}, G.~H.~A., {Groot}, P.~J., {Marsh}, T.~R., {et~al.} 2005,
  \bibinfo{title}{{SDSS J124058.03-015919.2: a new AM CVn star with a 37-min
  orbital period},} \mnras, 361, 487, \dodoi{10.1111/j.1365-2966.2005.09186.x}

\bibitem[{J. {Smak}(1981){Smak}}]{1981AcA....31..395S}
{Smak}, J. 1981, \bibinfo{title}{{On the emission lines from rotating gaseous
  disks},} Acta Astronomica, 31, 395

\bibitem[{J.-E. {Solheim}(2010){Solheim}}]{2010PASP..122.1133S}
{Solheim}, J.-E. 2010, \bibinfo{title}{{AM CVn Stars: Status and Challenges},}
  \pasp, 122, 1133, \dodoi{10.1086/656680}

\bibitem[{C.~C. {Steidel} {et~al.}(2004){Steidel}, {Shapley}, {Pettini},
  {Adelberger}, {Erb}, {Reddy}, \& {Hunt}}]{2004ApJ...604..534S}
{Steidel}, C.~C., {Shapley}, A.~E., {Pettini}, M., {et~al.} 2004,
  \bibinfo{title}{{A Survey of Star-forming Galaxies in the
  1.4<\raisebox{-0.5ex}\textasciitildeZ<\raisebox{-0.5ex}\textasciitilde 2.5
  Redshift Desert: Overview},} \apj, 604, 534, \dodoi{10.1086/381960}

\bibitem[{J. {van Roestel} {et~al.}(2021){van Roestel}, {Creter}, {Kupfer},
  {Szkody}, {Fuller}, {Green}, {Rich}, {Sepikas}, {Burdge}, {Caiazzo},
  {Mr{\'o}z}, {Prince}, {Duev}, {Graham}, {Shupe}, {Laher}, {Mahabal}, \&
  {Masci}}]{2021AJ....162..113V}
{van Roestel}, J., {Creter}, L., {Kupfer}, T., {et~al.} 2021,
  \bibinfo{title}{{A Systematic Search for Outbursting AM CVn Systems with the
  Zwicky Transient Facility},} \aj, 162, 113, \dodoi{10.3847/1538-3881/ac0622}

\bibitem[{J. {van Roestel} {et~al.}(2022){van Roestel}, {Kupfer}, {Green},
  {Wong}, {Bildsten}, {Burdge}, {Prince}, {Marsh}, {Szkody}, {Fremling},
  {Graham}, {Dhillon}, {Littlefair}, {Bellm}, {Coughlin}, {Duev}, {Goldstein},
  {Laher}, {Rusholme}, {Riddle}, {Dekany}, \& {Kulkarni}}]{2022MNRAS.512.5440V}
{van Roestel}, J., {Kupfer}, T., {Green}, M.~J., {et~al.} 2022,
  \bibinfo{title}{{Discovery and characterization of five new eclipsing AM CVn
  systems},} \mnras, 512, 5440, \dodoi{10.1093/mnras/stab2421}

\bibitem[{F. {Verbunt} \& S. {Rappaport}(1988){Verbunt} \&
  {Rappaport}}]{1988ApJ...332..193V}
{Verbunt}, F., \& {Rappaport}, S. 1988, \bibinfo{title}{{Mass Transfer
  Instabilities Due to Angular Momentum Flows in Close Binaries},} \apj, 332,
  193, \dodoi{10.1086/166645}

\bibitem[{P. {Virtanen} {et~al.}(2020){Virtanen}, {Gommers}, {Oliphant},
  {Haberland}, {Reddy}, {Cournapeau}, {Burovski}, {Peterson}, {Weckesser},
  {Bright}, {van der Walt}, {Brett}, {Wilson}, {Millman}, {Mayorov}, {Nelson},
  {Jones}, {Kern}, {Larson}, {Carey}, {Polat}, {Feng}, {Moore}, {VanderPlas},
  {Laxalde}, {Perktold}, {Cimrman}, {Henriksen}, {Quintero}, {Harris},
  {Archibald}, {Ribeiro}, {Pedregosa}, {van Mulbregt}, \& {SciPy 1. 0
  Contributors}}]{2020NaMet..17..261V}
{Virtanen}, P., {Gommers}, R., {Oliphant}, T.~E., {et~al.} 2020,
  \bibinfo{title}{{SciPy 1.0: fundamental algorithms for scientific computing
  in Python},} Nature Methods, 17, 261, \dodoi{10.1038/s41592-019-0686-2}

\bibitem[{T.~L.~S. {Wong} \& L. {Bildsten}(2021){Wong} \&
  {Bildsten}}]{2021ApJ...923..125W}
{Wong}, T. L.~S., \& {Bildsten}, L. 2021, \bibinfo{title}{{Mass Transfer and
  Stellar Evolution of the White Dwarfs in AM CVn Binaries},} \apj, 923, 125,
  \dodoi{10.3847/1538-4357/ac2b2a}

\bibitem[{J. {Wood} {et~al.}(1986){Wood}, {Horne}, {Berriman}, {Wade},
  {O'Donoghue}, \& {Warner}}]{1986MNRAS.219..629W}
{Wood}, J., {Horne}, K., {Berriman}, G., {et~al.} 1986,
  \bibinfo{title}{{High-speed photometry of the dwarf nova Z Cha in
  quiescence.},} \mnras, 219, 629, \dodoi{10.1093/mnras/219.3.629}

\end{thebibliography}
\bibliographystyle{aasjournalv7}

\end{document}